\documentclass{aa}
\usepackage{graphicx}
\usepackage{natbib}
\usepackage{amssymb}
\bibpunct{(}{)}{;}{a}{}{,} 

\begin{document}

\newcommand{\rmd}{{\mathrm d}}
\newcommand{\mm}{\mathrm}
\newcommand{\mi}{\mathit}

\title{X-Ray spectra from protons illuminating a neutron star}

\author{B.\ Deufel  \and C.P. Dullemond \and H.C.\ Spruit }

\offprints{bed@mpa-garching.mpg.de}

\institute{
       Max-Planck-Institut f\"ur Astrophysik, 
       Karl-Schwarzschildstr.~1, D-85740 Garching, Germany
          }

\date{Received / Accepted }

\abstract{We consider the interaction of a slowly rotating
  unmagnetized neutron star with a hot (ion supported, ADAF) accretion
  flow. The virialized protons of the ADAF penetrate into the neutron
  star atmosphere, heating a surface layer. Detailed calculations are
  presented of the equilibrium between heating by the protons,
  electron thermal conduction, bremsstrahlung and multiple Compton
  scattering in this layer. Its temperature is of the order 40-70 keV.
  Its optical depth increases with the incident proton energy flux,
  and is of the order unity for accretion at $10^{-2}$--$10^{-1}$ of
  the Eddington rate. At these rates, the X-ray spectrum produced by
  the layer has a hard tail extending to 100 keV, and is similar to
  the observed spectra of accreting neutron stars in their hard
  states. The steep gradient at the base of the heated layer gives
  rise to an excess of photons at the soft end of the spectrum
  (compared to a blackbody) through an `inverse photosphere effect'.
  The differences with respect to previous studies of similar problems
  are discussed, they are due mostly to a more accurate treatment of
  the proton penetration process and the vertical structure of the
  heated layer.  }

\maketitle

\section{Introduction}
\label{sec:one}

The spherical accretion of matter onto the surface of an unmagnetized
neutron star (NS) has attracted much attention since \cite{zel} [henceforce
ZS69] first addressed this problem more than thirty years ago.
As a model for the then newly discovered X-ray stars, these authors
considered a neutron star whose surface was heated by radially
infalling gas, and modeled this gas as consisting of freely falling
ions. A hot X-ray emitting layer is formed, the temperature of which
depends on the penetration depth of the protons into the atmosphere of
the star.  The model could reproduce the observed hard X-rays, but was
eclipsed by accretion disk models once it was realized that accretion
is not radial because of the angular momentum constraint.  The disk
model, however, only explains the sub--keV part of NS
spectra, somewhat in contradiction with observed spectra. In addition
to the optically thick disk, an additional source of hot gas thus had
to be found to produce the hard X-rays. This was proposed in the form of
the so-called two-temperature accretion flows
\citep{shapiro76,ichi77,rees82,narayan94,narayan95a}. In this form of
accretion, now called ADAF, the flow is geometrically thick, supported
by virialized protons, while the electrons stay at a lower temperature
around 100 keV due to their high radiative energy losses and the low
rate at which Coulomb interactions transfer accretion energy from the
ions to the electrons.
While it is not entirely clear at the moment how such a flow would
originate from the cool disk present at larger distances from the
star's surface, the observed hard X-ray component is often taken as an
indication of its existence. If accretion near the star takes
indeed place in the form of such a two-temperature plasma, the star's
surface is exposed to protons with energies around 50 MeV, and the
physics is very much like the model proposed by ZS69. It is therefore
interesting to pick up this line of theory and revisit
proton--injection models in context of NS--ADAF accretion.

In ZS69 the importance of the penetration depth of the accreted
protons for the outcoming spectrum was recognized. The penetration
length in terms of the amount of material required to stop the protons
was estimated and used as a model parameter. The injected energy of
the infalling protons per unit time and mass in the neutron star
atmosphere was then uniformly distributed in this stopping layer.  A
second parameter of their model was the accretion rate (or
luminosity). Their computed spectra essentially show a blackbody
spectrum with a high--energy tail due to Comptonization in the heated
atmosphere.

A more detailed numerical approach was presented by \cite{alme}.  They
introduced more physics into their model and also realized that the
proton deceleration depends on the atmospheric temperature. They
solved the atmospheric structure and the beam deceleration
simultaneously in a time dependent evolution of the model. But they
did not account for the strong dependence of proton deceleration
on the local proton velocity and used an estimate only for
the amount of material needed to stop the infalling protons. Their
resulting spectra again show a blackbody plus an additional high energy
tail.

\cite{bild92} were interested in the fate of the accreted CNO elements
in the neutron star atmosphere. They assumed an isothermal atmosphere
at a temperature defined by the accretion rate and the proton stopping
depth. Again a similar estimate for the penetration depth of the ions
was used.

The model of spherical accretion onto a neutron star was revisited by
\cite{turolla94}. They confirmed the existence of the formerly known
``cold'' solutions \citep[as in][]{alme}. In addition to this, they also
found ``hot'' solutions for the same luminosities.  But their work was
limited by the fact that the spectrum was just described in terms of a
mean photon energy.

\cite{zane98} confirmed the existence of the hot solutions with a much
more accurate treatment of the radiative processes including pair
processes. The emergent spectra of the hot solution should be
characterized by hard spectra peaking at $\approx 100$ keV.  Both
models use the stopping depth as a free input parameter. 
The effect of thermal conduction was not included in these calculations.

In continuation of this series of papers \cite{zampi95}
presented solutions for neutron stars accreting at low rates,
i.e. with luminosities $L_\mm{NS}=10^{-7}$ -- $ 10^{-3}
L_\mm{Edd}$.~They showed that the deviations from a blackbody spectrum
with the effective temperature of the neutron star increase with
decreasing accretion luminosities.

Proton illumination is not only restricted to the accretion of matter
onto the surface of a neutron star. \citet{spruit97}, \citet{spruit00}
and \citet{deufel00} have shown that this process might
also be of importance for accretion disks embedded in a hot corona.
With an improved version of the code used in \cite{deufel00} we
reconsider the accretion of matter onto the surface of a neutron star
(i.e. ion illumination of a neutron star surface). The code now
includes a better treatment of the radiative processes by including
thermal emission due to bremsstrahlung. We also allow for energy
redistribution due to electron thermal conduction within the
atmosphere.  The interaction between the accreted protons and the
neutron star atmosphere and the radiative processes are computed time
dependently in a one--dimensional, plane--parallel approximation.  The
density distribution through the atmosphere is found from hydrostatic
equilibrium including the pressure from the penetrating ions.

In the following we give a complete description of our model.  First
the heating of the electrons via Coulomb interactions is presented. We
show that the use of a stopping depth treated as a free parameter is
not sufficiently accurate for the calculation of the proton stopping. The
radiative processes are treated by solving the radiative transfer
equation instead of the Monte Carlo method used before. This has
advantages in terms of computation speed and the ability to treat
layers of arbitrary absorption optical depth.

\section{The Neutron star illumination model}
\label{sec:two}

\subsection{Proton stopping by Coulomb collisions}

We consider protons in the vicinity of a neutron star of mass
$M_\mm{*}$ and radius $R_\mm{*}$.  In ADAF accretion the typical
energy of such protons at the surface of a neutron star is the virial
temperature,

\begin{equation}
  \label{ep}
  E_\mm{p} = k\,T_\mm{vir} = \frac{G M_\mm{*} m_\mm{p}}{3 R_\mm{*}} =
  46 \,\frac{\mm{MeV}}{\mm{proton}} \left(
  \frac{M_\mm{*}}{M_{\odot}}\right ) \left( \frac{10
  \mm{km}}{R_\mm{*}}\right),
\end{equation}
where $G$ is the gravitational constant, $m_p$ the proton mass
and $k$ the Boltzmann constant.

If protons with such energies encounter a much a cooler and dense
medium, they will be stopped very efficiently by Coulomb interactions
with the cold electrons inside their Debye sphere. The stopping of a
fast particle in an ionized plasma was quantitatively discussed by
\cite{spitzer62}.  In a time $\rmd t$, such a proton with energy
$E_\mm{p}$ will loose an amount of energy $\rmd E$,

\begin{equation}
  \label{dep}
  \frac{d E}{d t}= -\varepsilon(E_\mm{p})\, n_\mm{e}\, v_\mm{p},
\end{equation}
where $n_\mm{e}$ is the electron density of the neutron star
atmosphere and $v_\mm{p}^2 \approx 3 k T / m_p$ is the proton 
velocity.~$\varepsilon(E_\mm{p})$ is called the stopping power of a plasma.
\cite{ryter70} show that it is given through Spitzers's formalism by

\begin{equation}
  \label{stpow}
  \varepsilon(E)=\frac{4 \pi e^4}{m_{\mm e}
    v_{\mm p}^2}\,\ln\Lambda\,[\psi(x)-x\psi'(x)].
\end{equation}
  
Here $\ln \Lambda=\ln[(3/2e^3)(k^3 T_{\mm e}^3/\pi
n_\mm{e})^{1/2}]$ is the Coulomb logarithm, $\sigma_T$ is the Thomson
cross-section, $\psi(x)$ and $\psi'(x)$ are the error function and its
derivative, and $x=(m_{\mm e} v_{\mm p}^2 / 2 k T_{\mm
  e})^{1/2} $ is the ratio of the incident proton velocity to
the thermal velocity of the electrons.  This formula holds for non
relativistic electron and proton temperatures.

Now we can express the energy loss of a proton moving at an angle
$\theta$ with respect to the vertical coordinate $z$, per unit of
vertical Thomson depth, $d\tau = n_\mm{e} \sigma_T \rmd z$, by

\begin{equation}
  \label{dedtau}
  \frac{d E}{d\tau_{\mm T}}= -\frac{1}{\sigma_T}\frac{4 \pi
  e^4}{m_{\mm e} v_{\mm p}^2\cos\theta}\,\ln\Lambda\,[\psi(x)-x\psi'(x)].
\end{equation}  

In a time-dependent evolution of the model we follow the protons
numerically through the neutron star atmosphere with temperature
$T(\tau)$ and density $n_\mm{e}(\tau)$.  We record the loss of kinetic
energy of the protons as a function of optical depth. This yields the
local time dependent heating rate $\Lambda_\mm{p}(\tau)$.  In this way
the stopping depth of the protons is obtained self consistently.

As mentioned in Sect. \ref{sec:one} the penetration depth
significantly influences the outcoming spectrum of the proton
illumination model. Usually a global estimate is used to calculate the
stopping depth of the protons in an atmosphere.  The estimate for the
stopping length $\lambda_\mm{s}$ is obtained following
\cite{spitzer62},

\begin{equation}
  \label{stpl}
  \lambda_\mm{s} = v_\mm{p} \,t_\mm{s}\;,
\end{equation}
where $t_\mm{s}$ is the slowing down time scale given by
\cite{spitzer62}, Eq.~(5-28). If we express the stopping length in
terms of Thomson optical depth, we get

\begin{equation}
  \label{stpth}
  \tau_\mm{s}=\frac{m_\mm{p}}{m_\mm{e}}\frac{2}{3 \ln \Lambda}
  \frac{(v_z/c)^4}{\psi - x\psi'},
\end{equation}
where $v_\mm{z}$ is the vertical component of the proton thermal
velocity and $c$ is the speed of light.  This expression is inaccurate
because of several factors. Most importantly the proton velocity
enters this formula as the fourth power. As the proton velocity
decreases, this factor drops rather fast and shortens the penetration
depth considerably. Counteracting is the function $\psi-x\psi'$. For
$x>2$ this function is unity. But for $0<x<2$ it is small, increasing
the stopping length but also reducing the energy loss [cf.
Eq.~(\ref{dedtau})].~Even the Coulomb logarithm should not be taken as
constant (as it is often done). In the wide range of electron
temperatures and densities of interest (as solutions from e.g.
\cite{turolla94} suggest) the Coulomb logarithm can easily vary by a
factor of two.

\begin{table}
\caption{\label{pdepth} Estimated stopping depth (est) and calculated
  stopping depth (num) 
    [according to Eq.~(\ref{dedtau})] in units of Thomson optical depth
    for a isothermal atmosphere with temperature $T_\mm{e}$ and uniform
    electron density $n_\mm{e} = 10^{22}$ cm$^{-3}$ for various proton
    energies.}
  \begin{tabular}{c|cccccc} 
    \hline
  \multicolumn{1}{c|}{} &
  \multicolumn{6}{c}{Temperature $T_\mm{e}$\,[keV]} \\ 
  \multicolumn{1}{c|}{$T_\mm{p}\, [MeV]$ } &
  \multicolumn{2}{c}{1} &
  \multicolumn{2}{c}{10} &
  \multicolumn{2}{c}{100} \\ \cline{2-7}
  & num & est & num & est & num & est \\
   \hline
23 & 0.2 & 0.8  & 0.4 & 0.8 & 6.8  & 7.8 \\
46 & 0.8 & 3.1  & 0.9 & 2.4 & 10.2 & 12.4 \\
92 & 3.1 & 12.4 & 2.6 & 8.9 & 15.9 & 21.6 \\
\hline
\end{tabular}
\end{table}

Table \ref{pdepth} gives some values for the estimated stopping depth
[from Eq.~(\ref{stpth})] and the calculated stopping depth [according to
Eq.~(\ref{dedtau})] by following a penetrating proton through an
isothermal electron layer until it has lost all of its energy in
excess with respect to the ambient electrons. The estimated stopping
depth is generally higher than the numerical result due to the reasons
mentioned above.  For low electron temperatures these discrepancies
are quite conspicuous.

The stopped protons accumulate at their stopping depth in the neutron
star atmosphere.  To maintain charge neutrality, an equal number of
electrons has to move to the same location. The ADAF electrons
entering the disk are stopped over a much short distance, compared
with the protons. To make them move to the location where the protons
come to rest in the disk, an electric field has to develop in the
neutron star atmosphere, such that the resulting electron current has
just this property. We assume that such an equilibrium actually
develops, on the grounds that any charge imbalance will quickly lead
to the buildup of a strong restoring electric field.

Eq.~(\ref{dedtau}) is only valid for non--relativistic conditions,
whose validity needs to be checked for the high proton temperatures in
an ADAF. Analytic expressions for relativistic temperatures exist only
in special cases.  \cite{step83} has derived such an expression for
the rate of transfer of energy between populations with relativistic
Maxwellian distributions in terms of an integral over the scattering
cross--section. In the case of hot protons heating cooler electrons
the Rutherford cross--section is the relevant cross--section. An
expression in closed form for the heating rate (in ergs cm$^{-3}$
sec$^{-1}$) is given by \cite{stepg83}, 

\begin{eqnarray}
  \label{eq:destep}
  q_\mm{Stp} & =  -\frac{3 m_\mm{e}}{2 m_\mm{p}} n_\mm{e}\,
  n_\mm{p}\, \sigma_\mm{T}\, c \times
  \frac{(kT_\mm{e}-kT_\mm{p})}{K_2(1/\theta_\mm{e})K_2(1/\theta_\mm{p})}\;
  \ln\Lambda
   \nonumber  \\ &    
  \times \left[\frac{2(\theta_\mm{e}+\theta_\mm{p})^2 + 1}
    {\theta_\mm{e}+\theta_\mm{p}}\;
    K_1\left(\frac{\theta_\mm{e}+\theta_\mm{p}}
      {\theta_\mm{e}\theta_\mm{p}}\right)
    + 2 K_0\left(\frac{\theta_\mm{e}+\theta_\mm{p}}
      {\theta_\mm{e}\theta_\mm{p}}\right)\right]\;,
\end{eqnarray}

where $\theta_\mm{e}=k T_\mm{e}/m_\mm{e} c^2$,$\,\theta_\mm{p}=k
T_\mm{p}/m_\mm{p} c^2$, and $K_0$, $K_1$, $K_2$ are Bessel functions.

The proton--electron heating rate according to the stopping formula
derived from Spitzer's theory is given by

\begin{equation}
  \label{hrate}
  q(v_\mm{p}) = n_\mm{e}\,n_\mm{p}\,v_\mm{p}\,\varepsilon(E_\mm{p})\;.
\end{equation}

\begin{figure}
\includegraphics[width=\hsize]{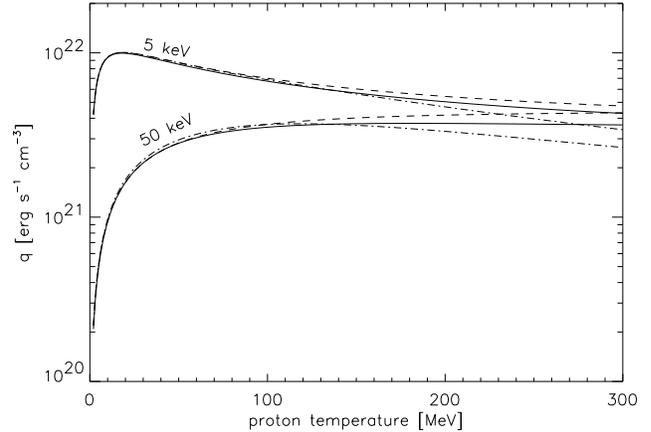}
\caption{Energy loss rates of a hot thermal distribution of protons in
a cooler ionized hydrogen plasma (with temperature indicated along the
curves). Solid line: fully relativistic expression of Stepney and
Guilbert (1983), dash-dotted line: the non relativistic expression from
Spitzer's (1962) theory. Dashed line: Spitzer's formula, but averaged
over a relativistic instead of a classical Maxwellian. Spitzer's
treatment is accurate to better than 5\% for proton temperatures $<
100$ MeV. }
\label{comp}
\end{figure}

By averaging Eq.~(\ref{hrate}) over a thermal velocity distribution we
obtain a rate that can be compared directly with Eq.~(15) from Stepney
\& Guilbert.  We do this averaging both with a classical Maxwell
distribution and a relativistic Maxwell distribution.  The results of
this comparison are shown in Fig. \ref{comp}. The heating rates are
almost indistinguishable up to proton temperatures of roughly 100 MeV.
At higher proton temperatures the deviations increase steadily. But
for the proton temperatures used in our simulation ($k T_\mm{p}<100$
MeV) Spitzer's formalism is accurate to better than 5\% compared to
Stepney's result.  Therefore we are confident that the usage of the
classical stopping formula following \cite{spitzer62} together with a
classical Maxwell distribution for the protons will yield the correct
results for to energy transfer from virialized protons to electrons
near a neutron star surface.

\subsection{The model atmosphere}
\label{sec:hydro}

We compute the density distribution through our plane--parallel one
dimensional model atmosphere from the equation of hydrostatic
equilibrium,

\begin{equation}
\label{press}
\frac{d P}{d \tau} = \frac{g}{\kappa_\mm{es}} + 
  \frac{\partial P_\mm{p}(\tau)}{\partial\tau}.
\end{equation}

Here $g = \frac{G M_{*} }{ R_{*}^2}$ is the gravitational
acceleration, which we treat as constant as the atmosphere is thin
compared to the neutron star radius, and $\kappa_\mm{es} = 0.40$
cm$^2$g$^{-1}$ is the electron scattering opacity. $\frac{\partial
P_\mm{p}}{\partial\tau}(\tau)$ denotes the force exerted by the
deceleration of the protons as a function of optical depth. The change
of momentum (exerted on the atmosphere) of an incoming proton with
respect to the vertical can be expressed by

\begin{equation}
  \label{dp}
  \Delta p_{\perp} = -m_\mm{p}\,\Delta v \cos \theta .
\end{equation}
The force $\partial P_\mm{p}/\partial\tau$ is evaluated by recording
the change of velocity $\Delta v$ of the incoming protons as a
function of optical depth. Since we only study cases with luminosities
below the Eddington limit we neglect the contribution of the radiative
pressure.

We solve our pressure profile by starting with $\tau_0$=0 at the top
of our atmosphere and integrate Eq.~(\ref{dp}) to the maximum
optical depth of our model atmosphere $\tau_\mm{max}$. We usually set
$\tau_\mm{max}=10$. At this optical depth the energy flux of the
protons is already negligible.

\subsection{Accretion from the ADAF}

The protons from the ADAF have a temperature of the order of the
virial temperature [Eq. (\ref{ep})]. As mentioned above we use a
Maxwell distribution for the proton velocities.  Additionally, the
ADAF has a rotation rate, which is somewhat smaller than the Keplerian
rate $\Omega_{\rm K}$ \citep[e.g.][]{narayan95a}. When computing the
penetration of the protons we take the Maxwellian velocity component
of the protons.  Further we take into account the component
of velocity tangential to the neutron star surface due to the ADAF
rotation.  Instead of using a detailed model of an ADAF in which these
velocity components can in principle be determined quantitatively, we
model it with a temperature $T=\xi T_{\rm vir}$ and tangential
velocity $v=\eta R\Omega_{\rm K}$, i.e. we use the parameters $\xi$
and $\eta$ to scale these quantities. Both the tangential and vertical
velocity components add to the energy deposited, but are not
equivalent since the thickness of the heated layer is smaller the more
the protons enter tangentially.

\subsection{The radiative transfer}
\label{sec:kees}

The radiative transfer equation appropriate for our model is
\begin{equation}\label{eq-transfer-eq-1}
\mu\frac{dI_{\mu,\nu}}{dz} = j^\mi{ff}_{\nu} + j^{c}_{\mu,\nu} 
- (\alpha^\mi{ff}_{\nu}+\alpha^{c}_{\nu}) I_{\mu,\nu}\;\;,
\end{equation}
where $I_{\mu,\nu}$ is the intensity as a function of frequency $\nu$
and photon angle $\mu=\cos(\theta)$. The coordinate $z$ is measured
from the top of the atmosphere where the density is virtually zero
($z=0$) downwards into the neutron star atmosphere until
$z=z_{\mathrm{max}}$ (corresponding to $\tau_\mm{max}$).

The bremsstrahlung absorption coefficient $\alpha^\mi{ff}_{\nu}$ is
given by
\begin{equation}\label{eq-brems-opac}
\alpha^\mi{ff}_\nu = 3.7\times 10^8\;n_\mm{e} n_\mm{i}\, \frac{1-e^{-h\nu/kT}}
{\nu^3\,\sqrt{T}} \,\bar g_\mi{ff}
\end{equation}
in units of cm$^{-1}$. The symbol $\bar g_\mi{ff}$ represents the
velocity averaged Gaunt factor for bremsstrahlung [see e.g.
\cite{rybicki79}]. The Compton extinction coefficient
$\alpha^{c}_{\nu}$ is simply
\begin{equation}\label{eq-compt-opac}
\alpha^{c}_\nu = n_e \sigma^c_\nu\;\;,
\end{equation}
where $\sigma^c_\nu$ is the electron cross section with Klein-Nishina
corrections. For the non-relativistic limit this equals the Thompson
cross section $\sigma_T$.

The bremsstrahlung emission $j^\mi{ff}_{\nu}$ is given by

\begin{equation}\label{eq-brems-emis}
j^\mi{ff}_{\nu} = 5.4\times 10^{-39} \;n_\mm{e} n_\mm{i} \,
\frac{e^{-h\nu/kT}}{\sqrt{T}}\,\bar g_\mi{ff} 
\end{equation}
in units of erg cm$^{-3}$ sec$^{-1}$ ster$^{-1}$. The Compton
`emissivity' (i.e.~radiation scattered to $\mu,\nu$ from some other
$\mu',\nu'$) is

\begin{equation}\label{eq-compton-emissivity-1}
j^{c}_{\mu,\nu} = n_\mm{e} \int \left(\frac{\nu}{\nu'}\right) I_{\mu',\nu'}
\,\sigma_c\,(\mu',\nu'\!\rightarrow\!\mu,\nu)\,d\mu'd\nu'\;\;.
\end{equation}

Here $\sigma_c(\mu',\nu'\!\rightarrow\!\mu,\nu)$ is the Compton
Scattering Kernel (CSK). An efficient algorithm for run-time
computation of the CSK is described by \cite{kershaw86} and \cite{kershaw87}.
Both $\sigma^c_\nu$ and $\sigma_c(\mu',\nu'\!\rightarrow\!\mu,\nu)$
depend on the electron temperature.

Whenever the temperatures of our solutions are low enough, $kT\ll
m_\mm{e}c^2$, we can simplify the Compton emissivity
Eq.~(\ref{eq-compton-emissivity-1}) without compromising accuracy. We
introduce the angle-averaged Compton cross section
$\sigma_c(\nu'\!\rightarrow\!\nu)$ defined as

\begin{equation}
\sigma_c(\nu'\!\rightarrow\!\nu) \equiv \frac{1}{2}
\int \sigma_c(\mu',\nu'\!\rightarrow\!\mu,\nu) d\mu'd\mu\;\;.
\end{equation}

A semi-analytic expression for $\sigma_c(\nu'\!\rightarrow\!\nu)$ is
given by \cite{pout96}.  The angle-averaged intensity $J_\nu$ is
defined as

\begin{equation}
J_\nu = \frac{1}{2}\int I_{\mu,\nu} d\mu\;\;.
\end{equation}

By replacing $\sigma_c(\mu',\nu'\!\rightarrow\!\mu,\nu)$ with its
average $\sigma_c(\nu'\!\rightarrow\!\nu)$ in
Eq.~(\ref{eq-compton-emissivity-1}) one obtains an approximative
expression for the Compton emissivity

\begin{equation}\label{eq-compton-emissivity-2}
j^{c}_{\nu} = n_\mm{e} \int \left(\frac{\nu}{\nu'}\right) J_{\nu'}
\sigma_c(\nu'\!\rightarrow\!\nu)d\nu'\;\;,
\end{equation}
which is independent of $\mu$, and which involves only a single integral
instead of the double integral of Eq.~(\ref{eq-compton-emissivity-1}).

Eqs.~(\ref{eq-transfer-eq-1},~\ref{eq-brems-opac},~\ref{eq-compt-opac},
~\ref{eq-brems-emis},
\ref{eq-compton-emissivity-1}/\ref{eq-compton-emissivity-2})
constitute the complete set of equations for the radiative transfer.
Since Eq.~(\ref{eq-transfer-eq-1}) and
Eq.~(\ref{eq-compton-emissivity-1}/\ref{eq-compton-emissivity-2}) are
mutually dependent, the solution cannot be found by direct evaluation.
To solve the system, we use a standard Lambda Iteration scheme [see
e.g. \cite{rutten99} and references therein; see also \cite{zane96},
\cite{pout96}].

We discretize frequency $\nu$ with equal spacing in $\log\nu$, and
photon angle $\mu=\cos(\theta)$ according to the roots of the Legendre
polynomials of order $n$. At the start of the procedure we choose
$j^c_{\mu,\nu}=0$, or any other initial guess that might be
appropriate. For each $\nu_i$ and $\mu_j$ we now integrate the
transfer equation (Eq.~\ref{eq-transfer-eq-1}) from $z=0$ to
$z=z_{\mathrm{max}}$ for $\mu_j>0$, or from $z=z_{\mathrm{max}}$ to
$z=0$ for $\mu_j<0$. At $z=0$ we impose zero inward flux as boundary
condition, while at $z=z_{\mathrm{max}}$ we choose a Planck function
at temperature $T_{\mathrm{BC}}$ as the starting value for the
integration. The temperature $T_{\mathrm{BC}}$ of this Planck function
is chosen such that:

\begin{equation}
\sigma T^4_{\mathrm{BC}} = 2\pi\int_0^1 \mu\, d\mu \int_0^\infty d\nu\;
I_{\mu,\nu}(z=z_{\mathrm{max}})
\end{equation}
so that at the bottom of the atmosphere the net flux is zero. 

After performing this operation for every $\nu_i$ and $\mu_j$, we can
evaluate the new $j^c_{\mu,\nu}$ by employing
Eq.~(\ref{eq-compton-emissivity-1}/\ref{eq-compton-emissivity-2}). This
new $j^c_{\mu,\nu}$ is inserted into Eq.~(\ref{eq-transfer-eq-1}) and
the integration of Eq.~(\ref{eq-transfer-eq-1}) is repeated along the
lines sketched above. After each iteration the relative difference of
the solution with the previous iteration step is computed. If this
relative difference drops below $10^{-3}$, we assume that the solution
has converged. Since in our solutions the scattering optical depths
remains always of the order of a few, we need not worry about the
well--known convergence problems of the lambda iteration procedure
sketched above.

Once the radiative transfer solution is obtained, the radiative
heating and cooling rates can be evaluated by flux differences. The
flux at each position $x$ is defined as

\begin{equation}
F(z) = 2\pi\int_{-1}^{+1} \mu\,d\mu \int_0^{\infty} d\nu \;I_{\mu,\nu}(x) \;\;.
\end{equation}

The net cooling rate is then

\begin{equation}
\label{net}
\Lambda_\mm{rad}^{-}(z) = \frac{dF(z)}{dz}\;\;.
\end{equation}

The radiative transfer is solved on a grid set up in terms of Thomson
optical depth when solving the pressure profile (Sect.
\ref{sec:hydro}).  This grid is then transferred to the vertical $z$
coordinates for the radiative transfer code by the usual

\begin{equation}
z(\tau) = \int_0^{\tau_\mm{max}}(n_\mm{e}(\tau) \sigma_T)^{-1} d \tau.
\end{equation}

Thus we immediately obtain the net cooling rates
$\Lambda_\mm{rad}^{-}(\tau)$ as a function of Thomson optical depth.

We also checked our solution for the importance of pair processes,
calculating the pair number densities according to \cite{zane98}. We
find that in our model the electron temperatures are too low and the
electron densities too high for pair production to become important.

\begin{figure*}
  \sidecaption \includegraphics[width=12cm]{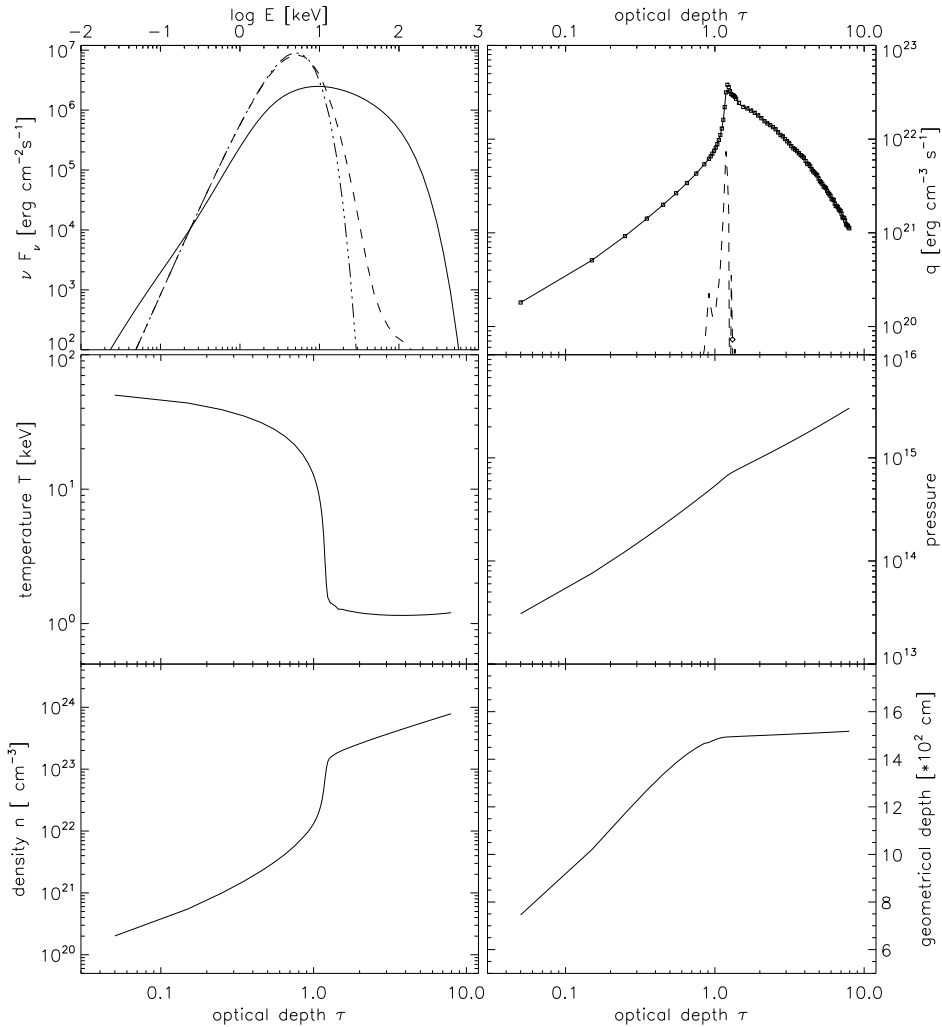} \caption{upper
    left: emergent spectrum (solid line), thermalized downward
    directed spectrum (dashed line) and blackbody input spectrum at
    the base (dotted -- dashed line); upper right panel: combined
    heating rates from proton heating and electron thermal
    conductivity (solid line) and radiative cooling rates due to
    Comptonization and bremsstrahlung (squares) - the dotted line
    shows the rates from electron conductivity alone; lower panels
    show from left to right and from top to bottom the electron
    temperature $T_\mm{e}$, pressure P, electron density $n_\mm{e}$
    and the geometrical depth of the layer $z$, for a solution with
    $F_\mm{p} = 1.6\times 10^{24}$ erg cm$^{-2}$ sec$^{-1}$, $T_\mm{p} =
    \frac{1}{2} T_\mm{vir}$ and $\Omega = \frac{1}{2}\Omega_\mm{K}$.}
    \label{view}
\end{figure*}

%
%
%
%
%
\begin{figure*}
\mbox{
\includegraphics[keepaspectratio=false,width=\columnwidth]{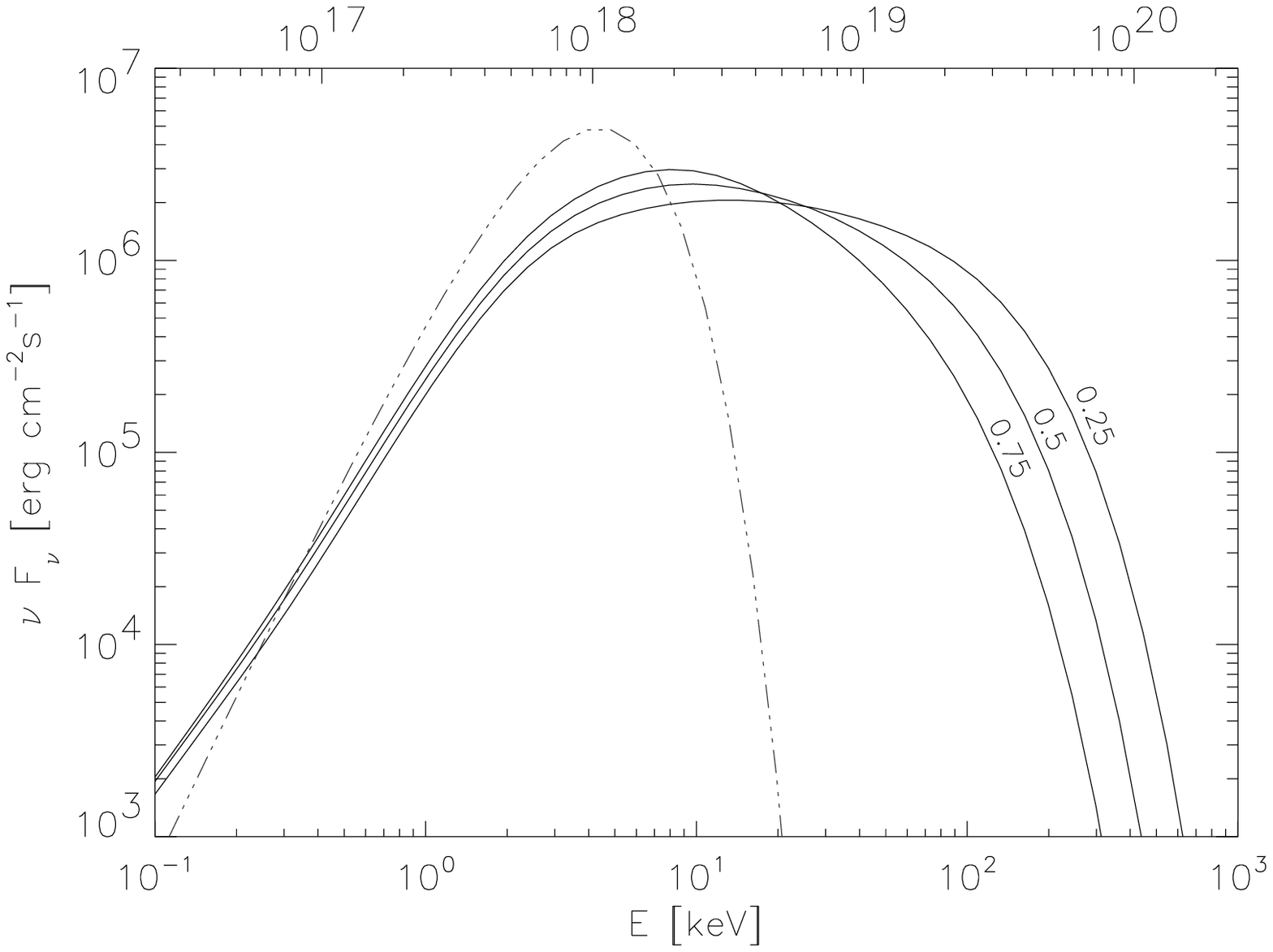}
\includegraphics[keepaspectratio=false,width=\columnwidth]{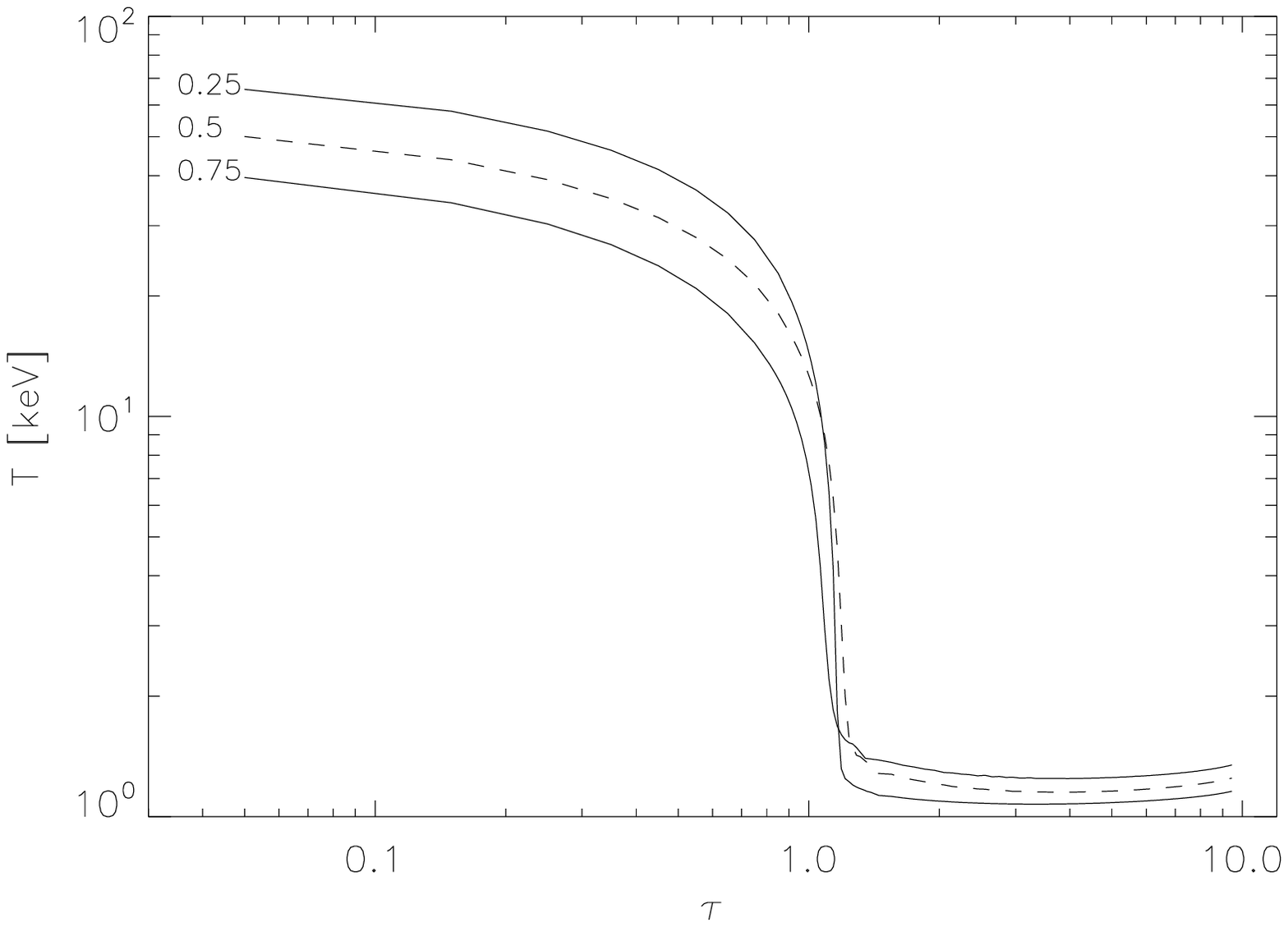}
}
\caption{Emergent model spectra and temperature profiles at a fixed
  proton energy flux $F_\mm{p} = 1.6\times 10^{24}$ erg cm$^{-2}$
  sec$^{-1}$ for various initial proton temperatures, $\xi\,T_\mm{vir}$,
  with $\xi$=0.25,\,0.5,\,0.75,\, and $\eta$=0.5. Higher proton
  temperatures produce lower electron temperatures in the hot part of
  the atmosphere and softer emergent spectra. The dotted dashed line
  shows the corresponding blackbody spectrum with the effective
  temperature of the layer, $T_\mm{eff} = (F_\mm{p}/\sigma)^{1/4}$.}

\label{varvir}
\end{figure*}
\begin{figure*}
\mbox{
\includegraphics[keepaspectratio=false,width=\columnwidth]{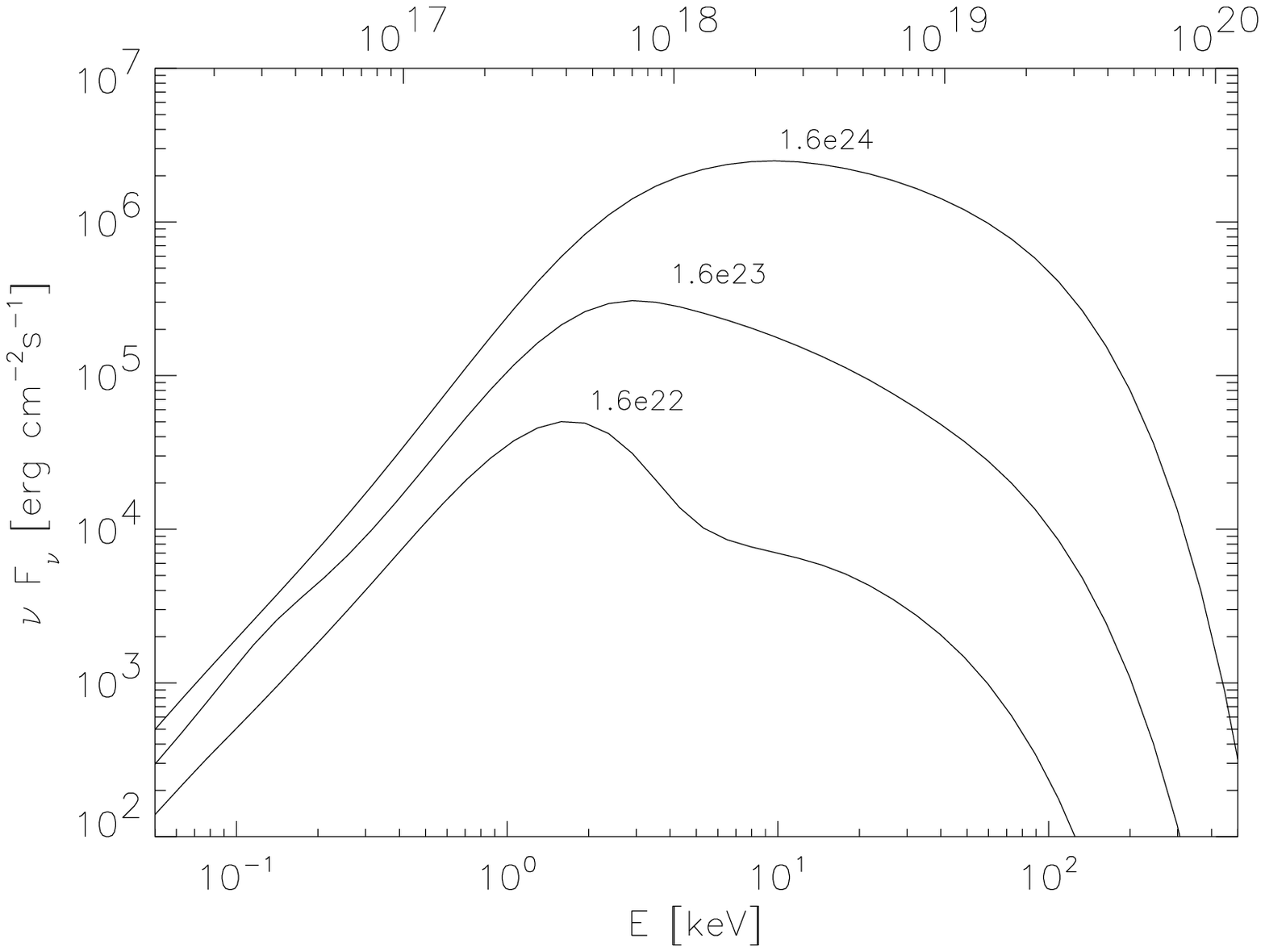}
\includegraphics[keepaspectratio=false,width=\columnwidth]{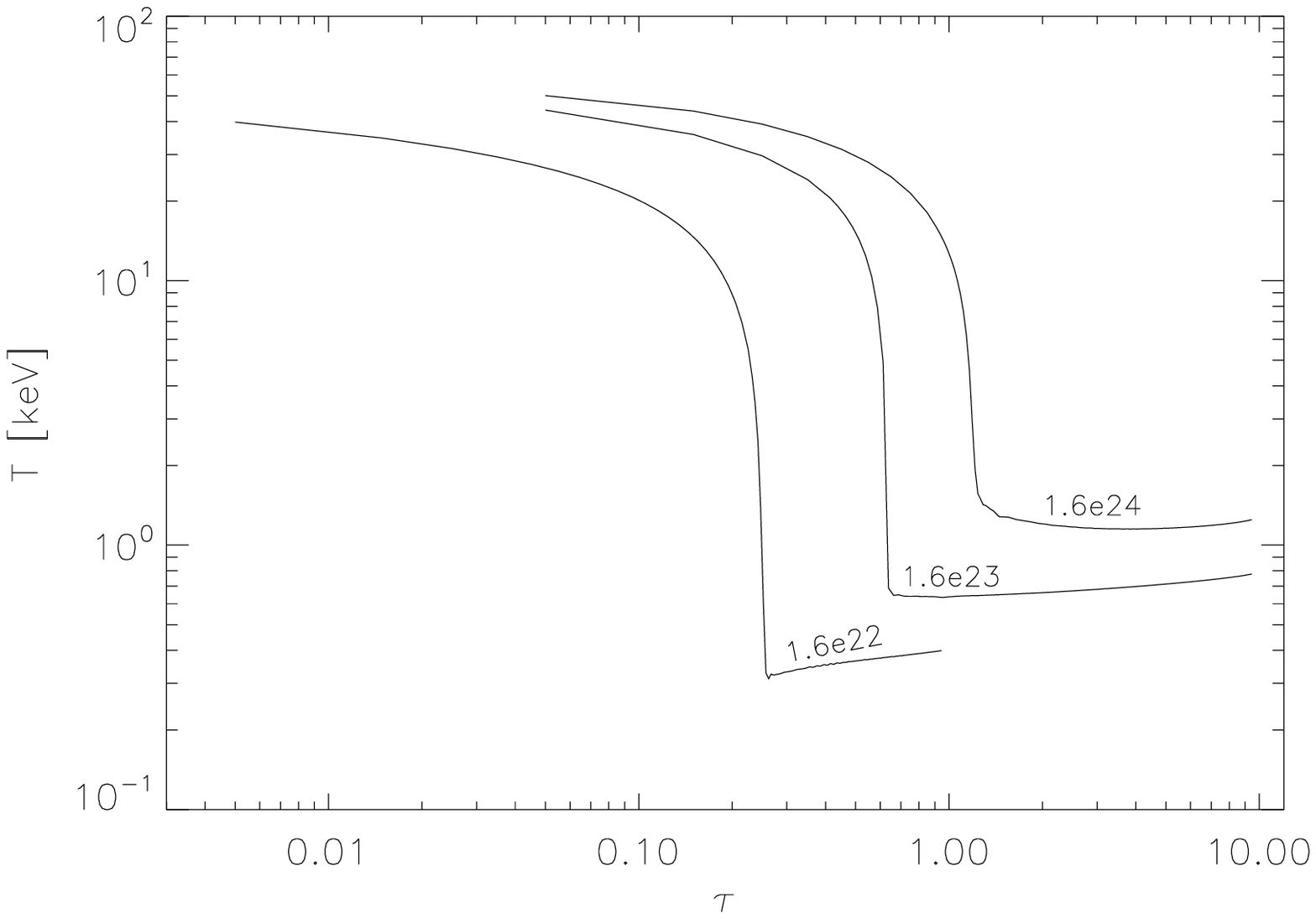}
}
\caption{Emergent spectra and temperature profiles for various values
  of the proton energy flux (in erg cm$^{-2}$ sec$^{-1}$, denoted by the
  numbers of the lines) at constant proton temperature, $\xi=0.5$.
  With decreasing energy flux the optical depth of the heated layer
  shrinks.}
\label{varflux}
\end{figure*}
\begin{figure*}
\mbox{
\includegraphics[keepaspectratio=false,width=\columnwidth]{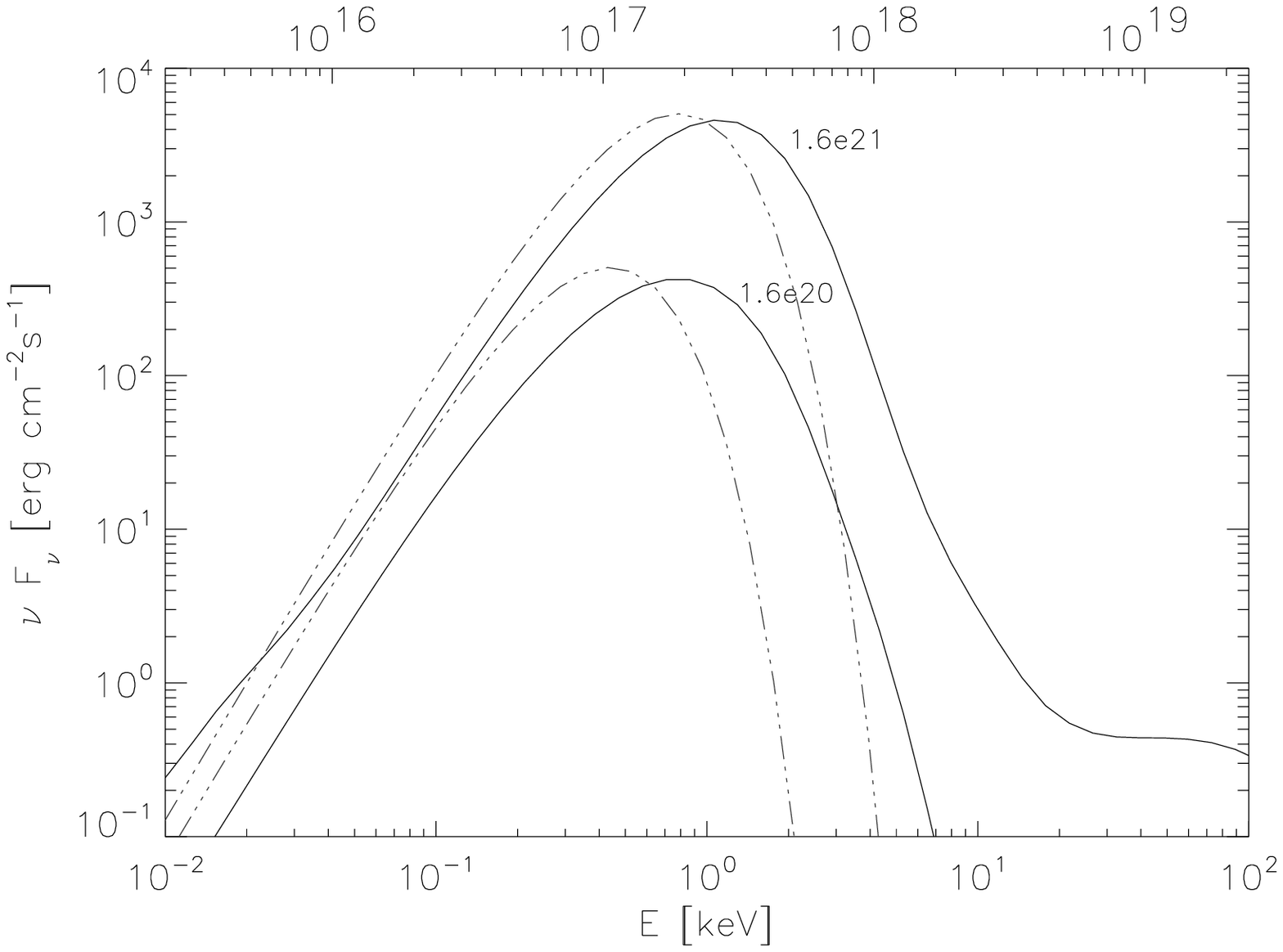}
\includegraphics[keepaspectratio=false,width=\columnwidth]{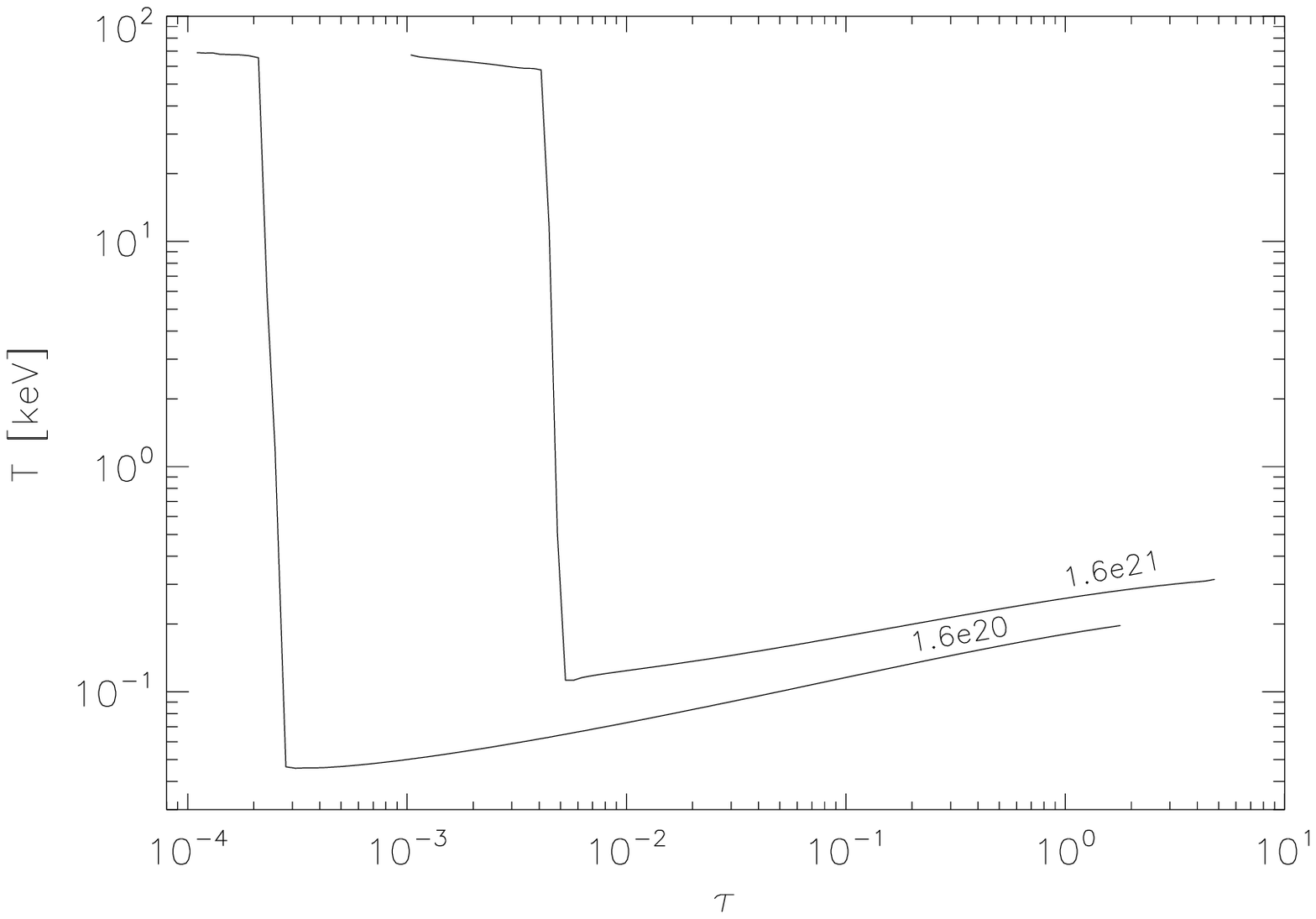}
}
\caption{Emergent spectra and temperature profiles at
  low proton energy flux (in erg cm$^{-2}$ sec$^{-1}$, denoted by the
  numbers of the lines) and constant proton temperature, $\xi=0.5$.
  The hot part is extremely thin in terms of Thomson optical depths.
  The dotted dashed lines show the corresponding blackbody spectra 
  with the effective temperature of the layer.}
\label{varflux1}
\end{figure*}

\subsection{Energy balance from heating and cooling}

We start our calculations with an isothermal atmosphere in hydrostatic
equilibrium according to Sect. \ref{sec:hydro}. For the initial
electron temperature we set $T_\mm{e}$ = 1 keV throughout the layer. After each
time step we obtain the heating rates $\Lambda_\mm{p}^+(\tau)$ from the
Coulomb interactions and the cooling rates $\Lambda_\mm{rad}^-(\tau)$ due to
the radiative processes bremsstrahlung and multiple Compton
scattering as a function of optical depth.
The time step of every cycle in our simulation is adjusted to the
shortest energy exchange time scale occurring in the calculation.

Additionally we include the energy redistribution due to electron
thermal conductivity. Generally the flow of heat per unit area, $Q$,
in the presence of a temperature gradient, is given by 
\citep[see e.g.][]{spitzer62}

\begin{equation}
\label{hcond}
Q \approx -\kappa\,\nabla T.
\end{equation}

The heat conductivity $\kappa$ for a fully ionized gas is given by

\begin{equation}
  \label{eq:kappa}
  \kappa \approx 1.85 \times 10^{-5}\,\frac{T^{5/2}}{\ln\Lambda} \;\;
  \mm{erg\; sec}^{-1}\mm{K}^{-1}\mm{cm}^{-1}.
\end{equation}

Thus the change of enthalpy $\partial w/\partial t$ due to electron
conductivity can be expressed by the divergence of the heat flux $Q$,

\begin{equation}
  \label{eq:dwcon} \frac{\partial w}{\partial t} = \rho\, c_p
  \,\frac{\partial T}{\partial t} =
  \nabla \cdot (\kappa \, \nabla T) =: \Lambda_\mm{cond}(\tau)
\end{equation}
where $c_p$ is the specific heat at constant pressure.

Now we can calculate the total change of enthalpy per time step,

\begin{equation}
  \label{ent} \frac{\Delta w(\tau)}{\Delta t} = \rho c_p \frac{\Delta
T(\tau)}{\Delta t} = \Lambda_\mm{p}^+(\tau) + \Lambda_\mm{rad}^-(\tau)
+ \Lambda_\mm{cond}(\tau).
\end{equation}

With Eq.~(\ref{ent}) we are able to calculate the change of temperature and
thus the new temperature profile after each time--step. With the new
temperature profile we can update the hydrostatic structure according
to Sect. \ref{sec:hydro}. We follow the simulation until the Coulomb
heating is balanced by the radiative cooling and energy redistribution
by conductivity and the temperature and density structure as well as
the spectra have reached a stationary state.

\section{Results of the model computations}
\label{sec:three}

For our model calculations we use fixed values for the mass $M_\mm{*}
= 1 M_{\sun}$ and the radius $R_\mm{*} = 10^6$ cm for the neutron star.
For the rotational angular velocity of the accreting material we use
$\eta = 0.5$.  The results depend on the proton energy flux per unit
area of the neutron star surface. In order to make the results
interpretable in terms of a total luminosity, we assume that 60\% of
the neutron star surface is involved in the accretion process. This
takes into account the approximate vertical extent of an ADAF
\citep{pop00}.

\subsection{Dependence on proton energy flux and temperature}

Fig. \ref{view} shows an equilibrium solution for a proton energy flux
$F_\mm{p} = 1.6\times 10^{24}$ erg cm$^{-2}$
sec$^{-1}$,\,$\xi=\eta=$0.5, yielding a luminosity of $L_\mm{*}=0.1
L_\mm{Edd}$.  The neutron star atmosphere is clearly divided into two
parts. A hot surface layer is separated by a sharp temperature
front from the much cooler bottom part. The maximum temperature in the hot
part of the atmosphere  is $T_\mm{e}\approx50$\,keV. This is
considerably hotter than previous calculations showed
\citep[e.g.][]{alme,turolla94} but also significantly cooler
than the ``hot'' solutions from \cite{turolla94} and \cite{zane98}. We
did not find any comparable hot solutions.

Fig. \ref{view} also shows that energy redistribution through electron
conductivity gets important at the transition zone from the hot to the
cool part.  The emergent model spectrum in Fig. \ref{view} illustrates
that the hot part acts as an effective Comptonization layer with an
optical thickness of order unity. The downward directed photons are
almost completely thermalized in the cold part. This justifies our
assumption of Sect. \ref{sec:kees} that the downward flux of photons
is converted into a black body flux of upward soft photons of
temperature $T_\mm{BC}$ at the base of our model since almost all the
thermalization has already taken place.

Fig. \ref{varvir} shows the emergent spectra and temperature profiles
through the atmosphere at a fixed proton energy flux level $F_\mm{p} =
1.6\times 10^{24}$ erg cm$^{-2}$ for various proton temperatures, $\xi
T_\mm{vir}$. The higher the initial proton temperature the lower is the
electron temperature in the hot part of the atmosphere and vice versa
for the cold part. The optical depth of the hot part is of order
unity in every case. As a consequence harder spectra are produced
with lower initial proton temperatures. This can be explained with the
decreasing penetration depth of the protons at smaller velocities: the
heating rates in the upper part of the atmosphere are increased and
thus the temperature is higher there. 

Fig. \ref{varflux} and Fig. \ref{varflux1} show the dependence of the
solution on the proton energy flux at constant proton temperature,
$\xi=0.5$. At low proton energy flux (Fig. \ref{varflux1}) the optical
depth of the hot part is small ($\tau_\mm{hot} < 10^{-2}$).
The solutions for these cases are comparable to those found by
\cite{zampi95}.  But the temperature of the hot part in our solutions
turns out to be much higher than in their solutions. This hot part
causes an (energetically unimportant) bump in the emergent spectra at
high frequencies due to optically thin bremsstrahlung emission in this
layer. As in \cite{zampi95} the emergent spectra are significantly
harder than the blackbody at the layer effective temperature,
$T_\mm{eff} = (F_\mm{p}/\sigma)^{1/4}$. We also see a harder spectrum
with respect to the blackbody of the layer effective temperature at
the lowest luminosity.

By further increasing the proton energy flux
the temperature jump moves inward to greater optical depth reaching
order unity at the highest flux levels. Thereby the temperature in the
hot part is roughly kept at a constant value in contrast to the
temperature of the cold part, which increases continously. At
$F_\mm{p} = 1.6\times 10^{24}$ erg cm$^{-2}$ sec$^{-1}$ the hot part
acts as an effective Comptonization layer for the cool thermal photons
from the interior. In addition the electron thermal conduction
smoothes the temperature profile and the shock is not as sharp as in the
low energy flux cases.

\subsection{A reverse photosphere effect}

\begin{figure}
\includegraphics[width=\hsize]{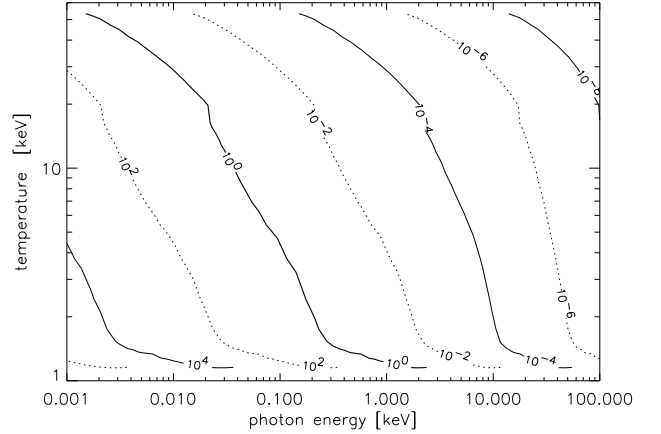}
\caption{Absorption optical depth as a function of photon energy and
  atmospheric temperature for an equilibrium solution with $\xi=0.5$
  and $\eta=0.5$. The line depicting $\tau_\mm{a}=1$ is emphasized.}
\label{odepth}
\end{figure}

Evident in our model spectra is an excess of soft photons (see e.g.
Figs.  \ref{view},\ref{varvir},\ref{varflux1}) with respect to a
blackbody with the effective temperature of the corresponding energy
flux. Though energetically unimportant it is an interesting feature of
our model spectra.  This soft photon excess is caused by the reverse
temperature profile of the atmosphere, i.e. the high temperatures are
above the cooler parts (in contrast to e.g. the solar photosphere).
Fig.  \ref{odepth} shows the absorption optical depth as a function of
the photon energy and electron temperature for an equilibrium solution
with $\xi=0.5$ and $\eta=0.5$.  The line for $\tau_\mm{a} = 1$ is
emphasized. For low photon energies $\tau_\mm{a} = 1$ is found high
in the neutron star atmosphere, where the temperature is high.
For high frequency photons the atmosphere gets more and more
transparent and $\tau_\mm{a} = 1$ occurs at lower temperatures.
Thus the high frequency photons decouple at lower temperatures and
propagate outwards. As a result there is an excess of low energy
photons, which decouple from the atmosphere at higher temperatures.

The same arguments explain the harder spectra at low proton energy
fluxes. Here the temperature profile in the atmosphere is dominated by
the increase of temperature with optical depth, as has already been
explained by \cite{zampi95}. Thereby the high energy photons decouple
in the deeper, hotter layers and propagate out freely. This
contribution to the high energy part of the spectrum causes the harder
spectrum compared to the blackbody of the layer effective temperature.

\section{Discussion and conclusions}
\label{sec:five}

The main conclusion from our work is that a neutron star surface
embedded in a hot ADAF--type accretion flow does not act like a simple
blackbody thermalizer. Instead the interaction with the hot ions
produces spectra with pronounced high energy, Comptonized tails.

We have considered the accretion of protons onto the surface of a
neutron star for a wide range of proton energy fluxes.  In our model
the stopping depth of the protons is obtained self consistently.  We
allow for the scaling of the proton temperature $T_\mm{p} = \xi
T_\mm{vir}$, which in our model controls the vertical velocity
component of the protons. Further we consider accretion from an ADAF,
where the protons have a tangential velocity component $v=\eta
\Omega_\mm{K}$.  By variation of the proton temperature and the proton
energy flux a range of X-ray spectra can be produced. Our
model computations show that the local proton energy flux determines the
optical depth of the hot part whereas the temperatures through the
atmosphere are determined by the initial proton temperatures.

\begin{figure}
  \includegraphics[width=\hsize]{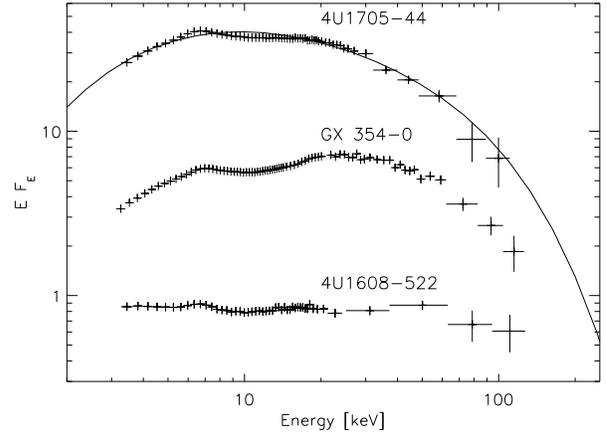}
\caption{Comparison of an emergent model spectrum with $\xi = 0.25$ and
  $\eta = 0.5$ with observed neutron star spectra in their hard
  states. The spectra are arbitrarily shifted in the vertical
  direction for clarity.}
\label{obs}
\end{figure}

Previous models considered spherical accretion onto the neutron star.
The free infall velocity of the protons was used in such models.  If
accretion takes place from an accretion disk or an ADAF, the vertical
velocity component is considerably smaller than than the free infall
velocity. This has important consequences on the emergent
spectra as our solutions suggest: at the same local energy flux a
decrease in the vertical proton velocity increases the temperature of
the hot part of the atmosphere, and X-ray spectra similar to those of
neutron stars in their hard states can be produced. Fig. \ref{obs}
shows a comparison of a model spectrum with observed spectra
of X-ray bursters in their low/hard states \citep[taken
from][]{gilv98}.

A question left unanswered by our studies is how the spectrum from the
surface is modified by the overlying accretion flow. E.g. an
overlying optically thick boundary layer might significantly
thermalize hard photons from the neutron star surface again. But we
can conclude that the input spectrum at the bottom of a boundary layer
is not Planckian at high proton energy fluxes. If the boundary layer
is optically thin the contribution to the observed hard spectral
component at high proton energy fluxes might be quite important.  A
discussion whether there is an optically thin boundary layer can be
found in e.g.  \cite{barret00}.  For low accretion rates the boundary
layer is indeed optically thin \citep[e.g.][]{king87}.

Detailed studies about the accretion disk boundary layer around a NS
have been performed by \cite{pop00}. They show that the accretion disk
near the NS is radially and vertically extended and that the angular
velocity $\Omega$ is reduced with respect to the Keplerian value. Such
a boundary layer naturally produces a proton illumination scenario as
described in our model. Their boundary layer is optically thin to
absorption. For high luminosities this region has a radial extent
larger than one stellar radius, i.e. the hard radiation from the NS
surface, produced by proton illumination, can easily propagate
outwards without being thermalized. The boundary layer even further
enhances Comptonization as the gas there is hot ($\gtrsim 10^8$~K).

Our results emphasize the importance of the neutron star surface for
the contribution to the hard spectral component in low mass X-ray
binaries (LMXBs).  A prerequisite for the occurrence of a hard spectral
component is a high local proton energy flux.  The local proton energy
flux onto the neutron star surface depends on both the mass accretion
rate and the size of the accretion belt on the neutron star. Our model
computations have local validity, i.e. if the involved accretion belt
is small, high proton energy fluxes on the neutron star surface can
even occur at accretion rates well beneath the Eddington limit.
Therefore our model does not exclude the possibility of hard spectral
components from the neutron star surface at low luminosities, as is
observed.

\begin{acknowledgements}
  We thank Roberto Turolla for the usage of the CSK subroutines. We are
  grateful to Marat Gilfanov for the neutron star spectra. This
  research has made use of data obtained through the HEASARC Online
  Service, provided by the NASA/GSFC. 
  
  This work was done in the research network ``Accretion onto black
  holes, compact stars and proto stars'' funded by the European
  Commission under contract number ERBFMRX-CT98-0195.
\end{acknowledgements}

\bibliographystyle{apj}
\bibliography{aamnem99,h2960}

\begin{thebibliography}{28}
\expandafter\ifx\csname natexlab\endcsname\relax\def\natexlab#1{#1}\fi

\bibitem[{{Alme} \& {Wilson}(1973)}]{alme}
{Alme}, M.~L. \& {Wilson}, J.~R. 1973, ApJ, 186, 1015

\bibitem[{{Barret} {et~al.}(2000){Barret}, {Olive}, {Boirin}, {Done},
  {Skinner}, \& {Grindlay}}]{barret00}
{Barret}, D., {Olive}, J.~F., {Boirin}, L., {Done}, C., {Skinner}, G.~K., \&
  {Grindlay}, J.~E. 2000, ApJ, 533, 329

\bibitem[{{Bildsten} {et~al.}(1992){Bildsten}, {Salpeter}, \&
  {Wasserman}}]{bild92}
{Bildsten}, L., {Salpeter}, E.~E., \& {Wasserman}, I. 1992, ApJ, 384, 143

\bibitem[{{Deufel} \& {Spruit}(2000)}]{deufel00}
{Deufel}, B. \& {Spruit}, H.~C. 2000, A\&A, 362, 1, paper I

\bibitem[{{Gilfanov} {et~al.}(1998){Gilfanov}, {Revnivtsev}, {Sunyaev}, \&
  {Churazov}}]{gilv98}
{Gilfanov}, M., {Revnivtsev}, M., {Sunyaev}, R., \& {Churazov}, E. 1998, Appl.
  Spectrosc., 338, L83

\bibitem[{{Ichimaru}(1977)}]{ichi77}
{Ichimaru}, S. 1977, ApJ, 214, 840

\bibitem[{Kershaw(1987)}]{kershaw87}
Kershaw, D. 1987, J. Quant. Spectrosc. Radiat. Transfer, 38, 347

\bibitem[{Kershaw {et~al.}(1986)Kershaw, Prasad, \& Beason}]{kershaw86}
Kershaw, D.~S., Prasad, M.~K., \& Beason, J.~D. 1986, J. Quant. Spectrosc.
  Radiat. Transfer, 36, 273

\bibitem[{{King} \& {Lasota}(1987)}]{king87}
{King}, A.~R. \& {Lasota}, J.~P. 1987, Appl. Spectrosc., 185, 155

\bibitem[{{Narayan} \& {Yi}(1994)}]{narayan94}
{Narayan}, R. \& {Yi}, I. 1994, ApJ Lett., 428, L13

\bibitem[{{Narayan} \& {Yi}(1995)}]{narayan95a}
---. 1995, ApJ, 444, 231

\bibitem[{{Popham} \& {Sunyaev}(2001)}]{pop00}
{Popham}, R. \& {Sunyaev}, R. 2001, ApJ, 547, 355

\bibitem[{{Poutanen} \& {Svensson}(1996)}]{pout96}
{Poutanen}, J. \& {Svensson}, R. 1996, ApJ, 470, 249+

\bibitem[{{Rees} {et~al.}(1982){Rees}, {Phinney}, {Begelman}, \&
  {Blandford}}]{rees82}
{Rees}, M.~J., {Phinney}, E.~S., {Begelman}, M.~C., \& {Blandford}, R.~D. 1982,
  Nat, 295, 17

\bibitem[{Rutten(1999)}]{rutten99}
Rutten, R. 1999, Radiative Transfer in Stellar Atmospheres,
  http://www.fys.ruu.nl/\~{}rutten/

\bibitem[{{Rybicki} \& {Lightman}(1979)}]{rybicki79}
{Rybicki}, G.~B. \& {Lightman}, A.~P. 1979, Radiative processes in astrophysics
  (New York: Wiley-Interscience)

\bibitem[{{Ryter} {et~al.}(1970){Ryter}, {Reeves}, {Gradsztajn}, \&
  {Audouze}}]{ryter70}
{Ryter}, C., {Reeves}, H., {Gradsztajn}, E., \& {Audouze}, J. 1970, A\&A, 8,
  389

\bibitem[{{Shapiro} {et~al.}(1976){Shapiro}, {Lightman}, \&
  {Eardley}}]{shapiro76}
{Shapiro}, S.~L., {Lightman}, A.~P., \& {Eardley}, D.~M. 1976, ApJ, 204, 187

\bibitem[{{Spitzer}(1962)}]{spitzer62}
{Spitzer}, L. 1962, Physics of fully ionized gases (New York: Wiley)

\bibitem[{{Spruit}(1997)}]{spruit97}
{Spruit}, H.~C. 1997, in Berlin Springer Verlag Lecture Notes in Physics, Vol.
  487, 67--76

\bibitem[{{Spruit} \& {Haardt}(2000)}]{spruit00}
{Spruit}, H.~C. \& {Haardt}, F. 2000, MNRAS, 315, 751

\bibitem[{{Stepney}(1983)}]{step83}
{Stepney}, S. 1983, MNRAS, 202, 467

\bibitem[{{Stepney} \& {Guilbert}(1983)}]{stepg83}
{Stepney}, S. \& {Guilbert}, P.~W. 1983, MNRAS, 204, 1269

\bibitem[{{Turolla} {et~al.}(1994){Turolla}, {Zampieri}, {Colpi}, \&
  {Treves}}]{turolla94}
{Turolla}, R., {Zampieri}, L., {Colpi}, M., \& {Treves}, A. 1994, ApJ Lett.,
  426, L35

\bibitem[{{Zampieri} {et~al.}(1995){Zampieri}, {Turolla}, {Zane}, \&
  {Treves}}]{zampi95}
{Zampieri}, L., {Turolla}, R., {Zane}, S., \& {Treves}, A. 1995, ApJ, 439, 849

\bibitem[{{Zane} {et~al.}(1996){Zane}, {Turolla}, {Nobili}, \& {Erna}}]{zane96}
{Zane}, S., {Turolla}, R., {Nobili}, L., \& {Erna}, M. 1996, ApJ, 466, 871+

\bibitem[{{Zane} {et~al.}(1998){Zane}, {Turolla}, \& {Treves}}]{zane98}
{Zane}, S., {Turolla}, R., \& {Treves}, A. 1998, ApJ, 501, 258+

\bibitem[{{Zel'dovich} \& {Shakura}(1969)}]{zel}
{Zel'dovich}, Y.~B. \& {Shakura}, N.~I. 1969, Soviet Astron.-AJ, 13, 175,
  {ZS69}

\end{thebibliography}

\end{document}